\documentclass[cits]{PoS}

\def\eg{\emph{e.g. }}
\def\ie{\emph{i.e. }}

\def\beq{\begin{equation}}
\def\eeq{\end{equation}}
\def\beqa{\begin{eqnarray}}
\def\eeqa{\end{eqnarray}}
\def\beg{\begin{lyxgreyedout}}
\def\eeg{\end{lyxgreyedout}}

\title{AdS/CFT as classical to quantum correspondence in a \emph{Virtual} Extra Dimension}

\ShortTitle{AdS/CFT from Elementary Cycles}

\FullConference{ICHEP2012, Melbourne, Australia}

\author{\speaker{Donatello Dolce}\\
	Camerino University, Italy \&  
         The University of Melbourne, 
Australia\\
   E-mail: \email{donatello.dolce@unicam.it}}

\abstract{The correspondence between classical extra dimensional geometry and quantum behavior, typical of the AdS/CFT, has a heuristic semiclassical interpretation in terms of undulatory mechanics and relativistic geometrodynamics. We note, in fact, that the quantum recurrence of ordinary particles enters into the equations of motions in formal duality with the extra dimensional dynamics of a Kaluza-Klein theory. The kinematics of the particle in a generic interaction scheme can be described as modulations of the spacetime recurrences and encoded in corresponding geometrodynamics. The quantization can be obtained semiclassically by means of boundary conditions, so that the interference of the classical paths with different windings numbers associated to the resulting recurrences turns out to be described by the ordinary Feynman Path Integral. This description applied to the Quark-Gluon-Plasma freeze-out yields basic aspects of AdS/QCD phenomenology. }

\FullConference{36th International Conference on High Energy Physics,\\
		July 4-11, 2012\\
		Melbourne, Australia}

\begin{document}


According to  Witten \cite{Witten:1998zw}, in AdS/CFT
 ``\emph{quantum phenomena [...] are encoded in classical geometry}'', without  however involving any explicit quantization condition. 
In this paper we investigate the origin of the ``classical to quantum'' correspondences of XD theories. It will be derived in terms: of Klein's original attempt to derive quantum mechanics from compact eXtra-dimension (XD) theory with Periodic Boundary Conditions (PBCs);  of Kaluza's and Nordstr\"om's  XD geometrical description of gauge interactions;  of de Broglie's assumption of spacetime recurrence associated to every particle.  This leads to a consistent semiclassical description of elementary particles quantum behaviors defined in \cite{Dolce:tune,Dolce:2009ce} and reviewed in \cite{Dolce:cycle}. Here we summarize results of \cite{Dolce:AdSCFT}.
  
In the atomistic description characterizing modern physics, every physical system is represented in terms of a set of elementary particles and their local retarded relativistic interactions. QM tells us,  through the Planck constant, that a spacetime recurrence of instantaneous periodicity $T^{\mu}=\{T_{t},\vec \lambda_{x}/c\}$  is associated  to every elementary particle of four-momentum $\bar p_{\mu}=\{\bar E /c,-\mathbf{\bar{p}}\}$.  As noted by de Broglie, 
the spacetime recurrence of a particle of mass $\bar M$  is fully characterized by the Compton time $T_{\tau} = h / \bar M c^{2}$, \ie the intrinsic periodicity of the proper time $\tau$,  or equivalently by the quantum recurrence $s\in (0,\lambda_{s}]$ of the worldline parameter $s= c \tau$, with $\lambda_{s} = c T_{\tau} $ (Compton length).  In a generic reference frame the spacetime recurrence resulting from this worldline periodicity is in fact described by the covariant relation $T_\tau \bar M c^2 \equiv T^{\mu} {\bar{p}}_{\mu} c \equiv h $. We have performed a Lorentz transformation $c T_\tau= c \gamma T_{t} - \gamma \vec \beta \cdot \vec \lambda_{x}$, $\bar E(\mathbf{\bar{p}}) = \gamma \bar M c^{2}$  and $\mathbf{\bar{p}} = \gamma \vec {\beta} \bar M c$.  This means that the classical-relativistic dynamics of a particle described by its 4-momentum $\bar p_{\mu} =\hbar \bar \omega_{\mu} / c $ can be equivalently encoded in retarded modulations of the corresponding local spacetime periodicity $T_{\mu}$. In undulatory mechanics elementary particles are described in terms of phasors or waves (``periodic phenomena'') in which the spacetime coordinates enter as angular variables. Their periodicities describe the kinematics of the particle though $\hbar$.  That is, every system in physics can be consistently described in terms of modulations of  elementary spacetime cycles with minimal topologies $\mathbb S^1$. 

We want to impose the intrinsic periodicity $T^{\mu}$ of elementary particles  as a constraint.   This represents a semiclassical quantization condition. A particle with intrinsic periodicity is similar to a ``particle in a box''. Through discrete Fourier transform the persistent periodicity $T^{\mu}$ directly implies a quantization of the conjugate spectrum $p_{n}^{\mu} = n \bar p^{\mu}$; $n$ is the quantum number associated to the topology $\mathbb S^{1}$. The quantization of the energy spectrum associated to the persistent time periodicity $T_{t}$ is the harmonic spectrum $E_{n} = n \bar E = n h / T_{t}$. A free bosonic particle can be therefore represented as a one dimensional bosonic string $\Phi(x)$ vibrating  in compact spacetime dimensions of length $T^{\mu}$ and Periodicity Boundary Conditions (PBCs --- denoted by the circle in $\oint$):
\begin{equation}
  {\mathcal{S}}^{\lambda_{s}} = \oint^{T^\mu}  d^4 x {\mathcal{L}}(\partial_\mu \Phi(x),\Phi(x)) =  \oint^{T'^{\mu} =\Lambda^\mu_\nu T^\nu} d^4 x' {\mathcal{L}}( \partial'_\mu \Phi'(x'),\Phi'(x'))\,.\label{generic:actin:comp4D}
\end{equation}
As known from string theory or XD theory, PBCs (or  combinations of Dirichlet and Neumann  BCs) minimizes the action at the boundary so that all the relativistic symmetries of (\ref{generic:actin:comp4D}) are preserved. This is a consequence of the fact that relativity fixes the differential structure of spacetime whereas the only requirement for the BCs is to fulfill the variational principle.  The expansion in harmonics of a field/string vibrating with persistent periodicity is $\Phi(x) = \sum_{n} \phi_n(x) = \sum_{n} A_{n} \exp[{-\frac{i}{\hbar} p_{n \mu} x^{\mu}}]$.  

To check the covariance we use a Lorentz transformation $x_\mu \rightarrow x'_\mu = \Lambda_\mu^\nu x_\nu$ as a generic transformation of variables in the free action (\ref{generic:actin:comp4D}) so that the transformed boundary of the resulting action yields a solution with transformed periodicity $T^\mu \rightarrow {T'}^\mu = \Lambda^\mu_\nu T^\nu$. This describes the four-momentum $\bar p_\mu \rightarrow  {{\bar p}'}_\mu = \Lambda_\mu^\nu \bar p_\nu$ of the free particle in the new frame,  according to $\bar p'_{\mu} c T'^{\mu} = h$. That is, $T^{\mu}$ transforms as a contravariant tangent 4-vector with the relativistic constraint induced by the underlying Minkowsky metric $\frac{1}{T^{2}_{\tau}} = \frac{1}{T_{\mu}} \frac{1}{T^{\mu}}$.   This is the geometric description in terms of  periodicities of the relativistic dispersion relation $\bar M c^{2} = \bar p_{\mu} \bar p^{\mu}$. Thus the harmonic energy spectrum of our system in a generic reference frame is $E_{n} (\bar \mathbf p) = n h / T_{t} (\bar \mathbf p) = n \sqrt{\bar \mathbf p^{2} c^{2} + \bar M^{2} c^{4}}$, which is the energy spectrum prescribed by ordinary second quantization (after normal ordering) for the single mode of  periodicity $T(\bar \mathbf p)$ of a free bosonic field. This is a first semiclassical correspondence with ordinary QFT. We also note a dualism to Kaluza-Klein (KK) theories. In the rest frame the proper time periodicity (compact worldline) describes a quantized rest energy spectrum, \ie a mass spectrum, $E_{n}(0)/c^{2} \equiv M_{n} = n \bar M = n h / \lambda_{s} c = n h / T_{\tau} c^{2} $, similar to a KK tower of compactification length $\lambda_{s}$.  
 
 As also noted by Einstein, a relativistic clock is ``phenomenon passing periodically through identical phases''. By assuming intrinsic periodicity every isolated particle can be therefore regarded as a reference clock. As in the Cs atomic clock whose reference ``tick'' of periods $10^{-10} s$ is fixed by an electronic energy gap, an isolated particle of energy $\bar E$ has  regular ``ticks'' of persistent periodicity $T_{t}$ that can be used  to define the unit of time. The so-called internal clock of an electron $T_{\tau} \sim 10^{-21} s$ has been observed in a recent experiment \cite{2008FoPh...38..659C}. The heavier the mass of the particle, the faster the periodicity  ($\bar E \sim 1$ TeV $\rightarrow T_{t} \sim 10^{{-27}} s$ ).  
  In a generic point $x=X$, a relativistic interaction of a particle can be characterized by the local retarded variations of four-momentum w.r.t. the free case $\bar{p}_{\mu}\rightarrow\bar{p}'_{\mu}(X)=e_{\mu}^{a}(x)|_{x=X}\bar{p}_{a}$. Through $\hbar$, interaction can be equivalently encoded by local retarded modulations of the internal clock of the particle $T^{\mu}\rightarrow T'^{\mu}(X)\sim e_{a}^{\mu}(x)|_{x=X}T^{a}$, that is by local ``stretching''  the compactification spacetime dimensions of (\ref{generic:actin:comp4D}). Therefore a generic interaction can be equivalently encoded in a locally deformed metric $\eta_{\mu\nu}\rightarrow g_{\mu\nu}(X)=[e_{\mu}^{a}(x)e_{\nu}^{b}(x)]|_{x=X}\eta_{ab}$. This description can be easily checked by using the local transformation of reference frame $dx_{\mu}\rightarrow dx'_{\mu}(X)=e_{\mu}^{a}(x)|_{x=X}dx_{a}$ as substitution of variables in the free action (\ref{generic:actin:comp4D}), \cite{Dolce:tune,Dolce:AdSCFT}. The resulting action with deformed metric $g_{\mu\nu}(X)$ describes a locally modulated solution of periodicity $T'^{\mu}(X)$: we pass from a free solution of persistent type  $\phi_{n}(x) \propto \exp[-\frac{i}{\hbar} p_{n \mu} x^{\mu}]  $  to the interacting solution of modulated type  $\phi'_{n}(x) \propto \exp[-\frac{i}{\hbar} \int^{x_{\mu}} d x'^{\mu} p_{n \mu}(x') ] $. Note also that in our formalism the kinematics of the interaction turns out to be equivalently encoded on the boundary, \emph{a la}  holographic principle.   Such a geometrodynamical description of generic interactions is of the same type of General Relativity. In a weak Newtonian interaction, the corresponding energy variation   $\bar{E}\rightarrow\bar{E}'\sim\left(1+{GM_{\odot}}/{|\mathbf{x}|c^2}\right)\bar{E}$  implies, through $\hbar$, a modulation of time periodicity $T_{t}\rightarrow T_{t}'\sim\left(1-{GM_{\odot}}/{|\mathbf{x}|c^2}\right)T_{t}$, \ie redshift and time dilatation. If we also consider the variation of momentum and the corresponding modulation of spatial periodicity, the resulting metric encoding the Newtonian interaction is actually the linearized Schwarzschild metric. Similar to Weyl's original proposal, gauge interaction can be obtained by considering local variations of flat reference frame $dx^{\mu}(x)\rightarrow dx'^{\mu}\sim dx^{\mu} - e dx^{a} \omega^{\;\mu}_{a}(x)$. Parametrizing by means of a vectorial field $\bar{A}_{\mu}(x) \equiv \omega_{\;\mu}^{a}(x)\bar{p}_{a}$, the resulting interaction scheme is actually $\bar{p}'_{\mu}(x)  \sim  \bar{p}_{\mu}- e\bar{A}_{\mu}(x)$, see \cite{Dolce:tune} for more details.

 Now we summarize the correspondence to ordinary relativistic quantum mechanics.  A vibrating string with modulated periodicity is the typical classical system that can be described locally in a Hilbert space.
 The modulated harmonics of such a string  form locally a complete set w.r.t  the corresponding local inner product $\left\langle \phi|\chi \right\rangle $. The harmonics defines  locally a Hilbert base $\left\langle x | \phi_{n}\right\rangle = \phi_{n}(x)$. Thus a modulated vibrating string, generic superposition of harmonics, is represented  by a generic Hilbert state $ \left| \phi \right\rangle = \sum a_{n} \left| \phi_{n} \right\rangle$. The non-homogeneous Hamiltonian $\mathcal H$ and momentum $\mathcal P_{i}$ operator are introduced as the operators associated to the 4-momentum spectrum of the locally modulate string: $\mathcal P_{\mu}\left |\phi_{n} \right\rangle = p_{n \mu}  \left|\phi_{n} \right\rangle$, where $\mathcal{P}_\mu = \{\mathcal H, - \mathcal{P}_{i}\}$.  From the modulated wave equation, the temporal and spatial evolution of every modulated harmonics satisfies $i \hbar \partial_{\mu} \phi_{n}(x) = p_{n}(x) \phi_{n}(x)$, thus the time evolution of our string with modulated periodicity represented by $\left\langle \phi|\chi \right\rangle $  is given by the ordinary Schr\"odinger equation $i \hbar \partial_{t} \left|\phi \right\rangle = \mathcal H \left|\phi \right\rangle$. Moreover, since we are assuming intrinsic periodicity, this classical-relativistic theory implicitly contains the ordinary commutation relations of QM. This can be seen by evaluating the expectation value of a total derivative $\partial_{x} F(x)$, and considering that the boundary terms of the integration by parts cancel each other owing the assumption of intrinsic periodicity. For generic Hilbert states we obtain: $i \hbar \partial_{x} F(x) = [F(x),\mathcal{P}]$ and $i \hbar  = [x,\mathcal{P}]$ for $F(x)=x$, \cite{Dolce:cycle,Dolce:tune,Dolce:AdSCFT,Dolce:2009ce}.  
 The correspondence with ordinary relativistic QM can also be seen from the fact that, remarkably, the classical evolution of such a relativistic vibrating string with all its modulated harmonics is described by the ordinary Feynman Path Integral (we are integrating over a sufficiently large number $N$ of spatial periods so that $V_{\mathrm{x}} = N \lambda_{x}$ is bigger than the interaction region) 
 \begin{equation}
\mathcal{Z}=\int_{V_{\mathrm{x}}} {\mathcal{D}\mathrm{x}}  \exp[{\frac{i}{\hbar}  \mathcal{S}(t_{f},t_{i})}]\,.\label{eq:Feynman:Path:Integral}\end{equation}
As usual, the $\mathcal S$ is the classical  action of the corresponding interaction scheme, with lagrangian $\mathcal{L} = \mathcal P x - \mathcal H $. This result has a very intuitive justification in the fact that in a cyclic geometry such as that associated to the topology $\mathbb S^{1}$, the classical evolution of $\phi(x)$ from an initial configuration to a final configuration  is given by the interference of all the possible classical paths with different windings numbers, \ie without relaxing the classical variational principle.  Thus the harmonics of the vibrating string/field $\phi(x)$ can be interpreted semiclassically as quantum excitations \cite{Dolce:cycle,Dolce:tune,Dolce:AdSCFT,Dolce:2009ce}. 

In this formalism it is straightforward to note that the cyclic worldline parameter enters into the equations in remarkable analogy with the cyclic XD of the KK theory \cite{Dolce:AdSCFT}. The solution $\phi$  can be in fact formally derived from a corresponding massless KK field by identifying the cyclic XD with a worldline parameter.   For this reason we address the cyclic worldline parameter $s$ as ``\emph{virtul} XD''.  If we in fact denote the XD and its compactification length with $s$ and $\lambda_{s}$ in a KK massless theory $dS^{2} = dx_{\mu} d x^{\mu} - d s^{2} \equiv 0$, and we identify the XD with the worldline parameter $s = c \tau$ we obtain our 4D theory  $d s^{2} = dx_{\mu} d x^{\mu} $ with cyclic worldline parameter of periodicity $\lambda_{s}$, and thus the spacetime periodicity $T^{\mu}$ by Lorentz transformation. In this case the quantized mass spectrum, the analogous of the KK tower, $M_{n} = n \bar M = n h / \lambda_{s} c$ is directly associated to the periodicity $\lambda_{s}$ of worldline parameter $s$ though discrete Fourier transform, whereas in the KK theory $M_{n}$ is obtained  indirectly through  the EoMs. That is, by assuming a VXD, the KK modes are \emph{virtual} in the sense that they are not 4D independent particles of mass $M_{n}$, they are the excitations of the same 4D elementary system. The quantized spectrum of a free boson is such that $p_{n \mu} x^\mu = M_{n} c s$ and thus we have the correspondence $\sum_n e^{-i p_{n \mu} x^\mu/\hbar}   \leftrightsquigarrow \sum_n e^{-i M_{n} c s/\hbar} $.  Such a collective description of the KK mode is typical of the holographic approach, where however a source field $\phi_{\Sigma}$ can be used as BCs to integrate out the heavy KK modes and achieve an effective description of the XD theory: $ {\mathcal{S}}^{5D}(s_{f},s_{i})\sim\mathcal{S}^{Holo}_{\Phi|_{\Sigma}=e\phi_{\Sigma}}(s_{f},s_{i})+\mathcal{O}(E^{eff}/\bar{M})\label{VXD:holo:appr} $.   This also means that, in analogy with the formalism of Light-Front-Quantization, the KK modes  form the base $\left| \phi_{n} \right\rangle$ of a Hilbert space, the evolution along the XD of a KK field is $i \hbar \partial_{s}\left| \phi \right\rangle = \mathcal M c \left| \phi \right\rangle $, where the mass operator $\mathcal M \left| \phi_{n} \right\rangle = M_{n }\left| \phi_{n} \right\rangle$ satisfies implicit commutation relations $[\mathcal M, s] = i \hbar$ owing the cyclic behavior of $s$. By means of the duality to XD theories, the generic interaction scheme $\bar{p}'_{\mu}(X)$ can be equivalently encoded in a corresponding deformed VXD metric $G_{MN} =\small{\left(\begin{array}{cc} g_{\mu\nu}& 0 \\ 0 & 1 \end{array}\right)}$ (this description should include dilatons or softwalls in the case of finite VXD). Under this dualism gauge interactions turn out to be encoded in a \emph{virtual} Kaluza metric.

By combining the correspondence  between classical cyclic dynamics and relativistic QM \cite{Dolce:2009ce}, the geometrodynamical formulation of interactions as modulation of spacetime periodicity \cite{Dolce:tune} and the dualism with XD theories in the holographic description, we obtain that the classical configurations of the modulated harmonic modes of $\phi$ in a curved XD background encodes the quantum behavior of the corresponding interaction scheme.  This correspondence for interacting particles is summarized by the following relation (with implicit source term \cite{Dolce:AdSCFT})
\begin{equation}
\int_{V_{\mathbf{x}}} {\mathcal{D}\mathbf{x}} \exp[{ \frac{i}{\hbar}\mathcal{S}'}] \leftrightsquigarrow 
\exp[{ \frac{i}{\hbar} \mathcal{S}^{Holo}_{\Phi|_{\Sigma}=e\phi_{\Sigma}}}]
\,.\label{VXD:QFT:corr}\end{equation}
Indeed, the description of physics in terms of elementary cycles pinpoints, at a semiclassical level, the  fundamental correspondence between classical XD geometry and 4D quantum behavior of AdS/CFT, \cite{Witten:1998zw}.
 To check this we consider the example of the Quark-Gluon-Plasma (QGP) freeze-out,  in which the classical dynamics of the interaction scheme are described by the Bjorken Hydrodynamical Model,  \cite{Magas:2003yp}. 
During the exponential freeze-out the 4-momentum of the QGP fields decays exponentially with the laboratory time, i.e. with the proper time: $\bar E \rightarrow \bar{E}(s) = e^{-ks/c}\bar{E}$.  In terms of QCD  thermodynamics,  $k$ represents the gradient of Newton's law of cooling. Thus, in the the massless approximation ($E \simeq c p$),  the 4-momentum of QGP during the freeze-out decreases conformally and exponentially  
$ \bar{p}_{\mu}\rightarrow\bar{p}'_{\mu}(s)\simeq e^{-ks/c}\bar{p}_{\mu}$. Equivalently, through the Planck constant, we have that  the spacetime periodicity has an exponential and conformal
dilatation
$
T^{\mu}\rightarrow T'^{\mu}(s)\simeq e^{ks/c}T^{\mu}\,.\label{eq:deform:4period:QGP}
$
According to our geometrodynamical description of interaction, this modulation of periodicity is therefore encoded in the substitution of variables
$
dx_{\mu}\rightarrow dx'_{\mu}(s)\simeq e^{-ks}dx_{\mu}\,.
$
The QGP freeze-out is thus encoded by the warped metric $ds^{2}=e^{-2ks/c}dx_{\mu}dx^{\mu}$. By treating the worldline parameter $s$  as a VXD, the exponential dilatation of the 
4-periodicity during the QGP freeze-out of massless fields  ($dS^{2}\equiv0$) can be equivalently encoded in the
\emph{virtual} AdS metric $
dS^{2}\simeq e^{-2ks/c}dx_{\mu}dx{}^{\mu}-ds^{2}\equiv0
$. 
The  energy of the QGP during the freeze-out is therefore parametrized, though the Planck constant, in terms of the time periodicity  $T_{t}(s) = e^{ks/c}/k = h / E(s)$. This is formally the conformal parameter $z(s) \equiv T_{t}(s)$ which in fact describes the inverse of the energy in ordinary AdS/CFT.  It varies from the initial state (\eg after the formation in a collider experiment) characterized by small time periodicity $T_{t}^{UV}=\frac{h}{\Lambda}=\frac{e^{ks^{UV}/c}}{k}$, to a  state characterized by large time periodicities $T_{t}^{IR}=\frac{h}{\mu}=\frac{e^{ks^{IR}/c}}{k}$. 
The massless approximation means infinite proper time periodicity, i.e. infinite VXD. Thus the AdS geometry encoding the freeze-out has no boundaries. Indeed, if we consider the propagation  of  a 5D gauge theory with 5D bulk coupling $g_{5}$ in an infinite VXD, the effective coupling of the corresponding 4D theory  behaves logarithmically w.r.t. the infrared scale
$
g^{2}\simeq\frac{g_{5}^{2}k}{\log\frac{\mu}{\Lambda}}
$. 
This reproduces the quantum behavior of the strong coupling constant as long as we suppose 
$
\frac{{1}}{{k}}\sim\frac{{N_{c}g_{5}^{2}}}{{12\pi^{2}}}
$, \cite{Pomarol:2000hp}. 
In agreement with the AdS/CFT dictionary, the classical dynamics associated to an infinite VXD actually encodes  the quantum behavior of a conformal theory. Indeed this has an intuitive justification in terms of undulatory mechanics and relativistic geometrodynamics. 
In our description, a massive system is characterized by a finite proper time periodicity $\lambda_{s}$. Thus, if we want to describe a QGP of massive fields we must assume a compact VXD. Through the holographic approach \cite{ArkaniHamed:2000ds} with small IR scale,  the classical configurations on this compact warped VXD is effectively described by  $\Pi^{Holo}(q^{2})\sim-\frac{q^{2}}{2kg_{5}^{2}}\log\frac{q^{2}}{\Lambda^{2}}$. This approximately matches the  two-point function of QCD and the asymptotic freedom
$
\frac{1}{e_{eff}^{2}(q)}\simeq\frac{1}{e^{2}}-\frac{{N_{c}}}{{12\pi^{2}}}\log\frac{q}{\Lambda}. 
$ 
This quantum behavior has been obtained without imposing  any
explicit quantization except BCs. We note that in a consistent description of the massive case we must also abandon the conformal behavior between the temporal and spatial components ($T_{t} \neq \lambda_{x} /c$). This means the AdS metric must be consistently deformed, for instance, by introducing dilatons in the metric or ``soft-walls''.  We know that these geometries reproduces a realistic hadronic spectrum. Similarly to Veneziano's original idea of strings, in our description the hadrons are indeed energy (quantum) excitations, \ie \emph{virtual} KK modes, of the same fundamental string vibrating with characteristic compact worldline parameter and deformed spacetime encoding the interaction.  Such a geometrodynamical description of the masses is relevant to understand the  gauge symmetry breaking and thus of the Higgs mechanism \cite{Dolce:tune,Dolce:AdSCFT}. 



In AdS/CFT \emph{quantum behavior [...] are encoded in classical geometry} \cite{Witten:1998zw}. 
We conclude that this central aspect of AdS/CFT has a heuristic semiclassical  justification in terms  of undulatory mechanics and relativistic geometrodynamics \cite{Dolce:AdSCFT}.  
The quantization is equivalently obtained semiclassically by means of PBCs, in analogy with a ``particle in a box'', or Light-Front-Quantization. 
Every quantum particle is represented as a classical string vibrating in spacetime(minimal topology $\mathbb S^{1}$) whose harmonic energy levels are the quantum excitations of the system, as proven by the formal correspondence to ordinary QFT \cite{Dolce:2009ce}. 
Such a pure 4D  description of elementary particles has an explicit dualism with XD theories. The cyclic worldline parameter of the theory enters into the equations in formal analogy with the XD of a KK theory.  The KK modes turn out to encode quantum excitations of the same 4D system. In analogy with general relativity, the local spacetime modulations of periodicity encoding a given interaction scheme (i.e. local variations of four-momentum) can be equivalently described in terms of spacetime geometrodynamics. As show in \cite{Dolce:tune} such a description also yields ordinary gauge interactions, similarly to original Kaluza's and Weyl's proposal.  
By combining all these correspondences of the dynamics in modulated compact spacetime 
we have inferred semiclassically that the classical configurations in a deformed  VXD geometry reproduces the quantum behavior of a corresponding interaction scheme \cite{Dolce:AdSCFT}. 
Though AdS/CFT has a very rich phenomenology only partially investigated here and the validity of our approach is semiclassical, our description confirms the classical to quantum correspondence noted in \cite{Witten:1998zw},
and the application to the QGP  freeze-out  yields analogies to AdS/QCD.



%

\end{document}